\documentclass[sigconf]{acmart}

\AtBeginDocument{%
  }

\setcopyright{acmlicensed}
\copyrightyear{2024}
\acmYear{2024}
\acmDOI{XXXXXXX.XXXXXXX}

\acmConference[Conference acronym 'ICAIF 2024]{Make sure to enter the correct
  conference title from your rights confirmation emai}{Nov 14--17,
  2024}{Brooklyn, NY}
\acmISBN{978-1-4503-XXXX-X/18/06}

\usepackage{times}
\usepackage{soul}
\usepackage{url}
\usepackage[utf8]{inputenc}
\usepackage{caption}
\usepackage{subcaption}
\usepackage{graphicx}
\usepackage{amsmath}
\usepackage{amsthm}
\usepackage{booktabs}
\usepackage{algorithm}
\usepackage{algorithmic}
\usepackage[switch]{lineno}
\DeclareMathOperator{\E}{\mathbb{E}}
\DeclareMathOperator*{\argmax}{arg\,max} 
\begin{document}

\title{Hedging and Pricing Structured Products Featuring Multiple Underlying Assets}

\author{Anil Sharma}
\email{anil.sharma2@in.ey.com}
\orcid{https://orcid.org/0000-0003-0100-6140}
\affiliation{%
  \institution{Ernst \& Young}
  \country{India}
}
\author{Freeman Chen}
\email{freeman.chen@ca.ey.com}
\affiliation{%
  \institution{Ernst \& Young LLP}
  \country{Canada}
}

\author{Jaesun Noh}
\email{jaesun.noh@ca.ey.com}
\affiliation{%
  \institution{Ernst \& Young LLP}
  \country{Canada}
}

\author{Julio DeJesus}
\email{julio.dejesus@ca.ey.com}
\affiliation{%
  \institution{Ernst \& Young LLP}
  \country{Canada}
}

\author{Mario Schlener}
\email{mario.schlener@ca.ey.com}
\affiliation{%
  \institution{Ernst \& Young LLP}
  \country{Canada}
}









\begin{abstract}
    Hedging a portfolio containing autocallable notes presents unique challenges due to the complex risk profile of these financial instruments. 
    In addition to hedging, pricing these notes, particularly when multiple underlying assets are involved, adds another layer of complexity.    
    Pricing autocallable notes involves intricate considerations of various risk factors, including underlying assets, interest rates, and volatility. Traditional pricing methods, such as sample-based Monte Carlo simulations, are often time-consuming and impractical for long maturities, particularly when there are multiple underlying assets. In this paper, we explore autocallable structured notes with three underlying assets and 
    proposes a machine learning-based pricing method that significantly improves efficiency, computing prices 250 times faster than traditional Monte Carlo simulation based method.
    
    Additionally, we introduce a Distributional Reinforcement Learning (RL) algorithm to hedge a portfolio containing an autocallable structured note.
    Our distributional RL based hedging strategy provides better $PnL$ compared to traditional Delta-neutral and Delta-Gamma neutral hedging strategies. The $VaR$ $5\%$ ($PnL$ value) of our RL agent based hedging is $33.95$, significantly outperforming both the Delta neutral strategy, which has a $VaR$ $5\%$ of $-0.04$, and the Delta-Gamma neutral strategy, which has a $VaR$ $5\%$ of $13.05$. It also provides the hedging action with better left tail $PnL$, such as $95\%$ and $99\%$ value-at-risk ($VaR$) and conditional value-at-risk ($CVaR$), highlighting its potential for front-office hedging and risk management. 

\end{abstract}





\received{20 February 2007}
\received[revised]{12 March 2009}
\received[accepted]{5 June 2009}

\maketitle

\section{Introduction}
\label{sec:intro}

Autocallable notes are complex structured financial products that offer investors potential returns linked to the performance of an underlying asset, such as a stock or index. These notes have an embedded "autocall" feature, meaning they can be automatically redeemed before maturity if the underlying asset meets certain predefined conditions on specified observation dates. The complexity of autocallable notes stems from their intricate components, such as barrier levels, coupon payments, and conditions for early call. This complexity poses significant challenges for both pricing and hedging, necessitating advanced financial models and a thorough understanding of market dynamics. The difficulty increases substantially when autocallable notes involve multiple underlying assets, such as a combination of different stocks or indices. 
 The inclusion of potential coupon and call features complicates the price profile over time.
Traditional methods for pricing, such as sample-based Monte Carlo pricers, are often very time-consuming, especially for long maturities involving multiple underlying assets, making them impractical in many cases.
This inefficiency becomes especially problematic in applications such as Reinforcement Learning based hedging and XVA calculations. In Reinforcement Learning, where millions of scenarios must be processed for effective hedging, traditional pricers are too slow to be practical. Similarly, for XVA calculations, which require the pricing of instruments hundreds of thousands of times, using a faster pricer can significantly reduce computation time and enhance efficiency.



Utilizing machine learning (ML) to approximate the original pricing model offers substantial efficiency gains~\cite{ML_MC_1_alexei,ML_MC_2_Mcghee2018AnAN,ML_MC_3_neufeld2022}. In this direction, we propose a machine learning based approximation which enhances efficiency, expediting the pricing process and significantly reducing computational time compared to the traditional methods. For instance, in a case of autocallable notes with three underlying indexes, our ML approximator computes the pricing 250 times faster than the Monte Carlo based pricer. Moreover, the execution time of the ML approximator remains constant regardless of the original pricer's complexity, resulting in significant efficiency enhancements. 

In addition to the pricing, hedging a portfolio containing autocallable notes with multiple underlying assets is also crucial due to the complex risk profile of these financial instruments. 
Effective hedging strategies are necessary to ensure that the portfolio is protected against adverse price movements and unexpected changes in asset correlations. Additionally, given the intricacies of pricing and managing such notes, sophisticated hedging techniques are required to optimize risk management and enhance portfolio performance against adverse movements in the portfolio Gamma or index prices. 

Employing RL techniques to hedge such portfolios allows for the dynamic adjustment of hedging positions~\cite{kolm_ritter}, ensuring that the portfolio remains well-protected against market fluctuations while capitalizing on the benefits of multiple underlying assets. Other researchers have also demonstrated that RL is an attractive alternative to traditional hedging strategies based on the performance and Profit and Loss ($PnL$) distribution as compared to the baseline Delta neutral and Delta-Gamma neutral strategies~\cite{2q_Cao_2020,kolm_ritter,cao2023_gamma_vega}. Recently, \cite{autocall_hedge_mc} proposed a method for pricing autocallable notes and employing a Delta neutral strategy for hedging. They emphasized that hedging autocallable notes is complex due to their intricate structure. Similarly, we investigate autocallable structured note hedging using RL which outperforms the traditional hedging strategies (Delta neutral and Delta-Gamma neutral). In particular, the structured products, which derive their value from multiple underlying assets, can offer diversified risk exposure and enhanced return potential. Employing RL techniques to hedge such portfolios allows for the dynamic adjustment of hedging positions, ensuring that the portfolio remains well-protected against market fluctuations while capitalizing on the benefits of multiple underlying assets.

In our approach, we use Distributed Distributional DDPG (D4PG) algorithm with Quantile Regression (QR) to learn an optimal policy for hedging. This distributional RL enables a more nuanced understanding of uncertainty and risk in the learning process. Specifically, we use an American option as hedging instruments and the trained RL agent selects an action which quantifies the amount of hedging that needs to be performed. We compare the $PnL$ distribution, Value at Risk ($VaR$), and Conditional Value at Risk ($CVaR$) of different hedging strategies and show that the RL algorithm not only reduces $95\%VaR$, but also makes the $PnL$ distribution more symmetric and retains positive returns. By employing our advanced pricing model, we have achieved very fast results, significantly reducing the time required for pricing and eventually hedging. Furthermore, our RL-based hedging strategy has demonstrated superior performance compared to traditional Gamma hedging, offering more effective risk mitigation. This approach not only simplifies the pricing and hedging processes but also enhances the overall efficiency and robustness of portfolio management in the face of dynamic market conditions. The prices for the underlying asset are generated using Geometric Brownian motion (GBM) model.

Our specific contributions are the following: 
\begin{enumerate}
    \item We propose a machine learning-based option pricer for autocallable structured notes that computes prices 250 times faster than traditional Monte Carlo methods.
    \item We use a distributional RL based method to hedge a portfolio containing one short autocallable note under multiple underlying assets. We conduct a thorough analysis and introduce a novel objective function which helps to learn a generalized policy that beats traditional hedging strategies. 
    \item We compare with traditional hedging strategies, including Delta neutral and Delta-Gamma neutral and show that RL hedging outperforms these traditional methods.
    
\end{enumerate}


\section{Related Works}
\label{sec:related}

In this section, we review the literature on option pricing and hedging strategies specifically for exotic options and structured products.

\subsection{Option Pricing}
Monte Carlo simulation is the most commonly used technique in derivatives pricing; however, it is often computationally intensive. To accelerate the calculation process, machine learning (ML), deep learning (DL), and neural networks (NN) are extensively employed to approximate the original pricing function, which is also referred to as model-free pricing \cite{ML_MC_1_alexei,ML_MC_2_Mcghee2018AnAN,ML_MC_3_neufeld2022}. In finance, regulators require that these approximations be explainable and predictable. Unfortunately, NN methods typically do not meet these criteria. Several alternative methods satisfy the explainability and predictability conditions while also being efficient, such as Fourier and Chebyshev series expansions, Image rendering methods based on regular and stochastic sampling, and Tensor decomposition methods~\cite{ML_MC_4_antonov2021}.

In this paper, we apply the Chebyshev Tensor method as our model-free machine learning pricing approach. This technique uses Chebyshev polynomials and tensor decomposition to approximate pricing functions efficiently and accurately. By leveraging these mathematical tools, we achieve a stable, computationally efficient method that meets regulatory requirements for explainability and predictability. Previous studies have shown the effectiveness of Chebyshev Tensor methods in financial applications such as FRTB compliance and market risk assessment~\cite{ML_MC_5_zeron2023frtb,ML_MC_6_maran2021chebyshev,ML_MC_7_ruiz2021machine,ML_MC_8_zeron2021denting}, and our work aims to further validate its utility in derivative pricing.

\subsection{Hedging}
In recent years, Reinforcement Learning (RL) methods have been extensively applied in finance for a variety of purposes, including optimal trade execution~\cite{zhang2023generalizable_optimaltrade}, credit pricing~\cite{khraishi2022offline_creditpricing_icaif22}, market making~\cite{ganesh2019reinforcement_marketmaking}, learning exercise policies for American options~\cite{pmlr-v5-li09d_earlyexerciseRL}, and optimal hedging~\cite{2q_Cao_2020,kolm_ritter,deeper_hedging_Gao_2023,murray2022deep_icaif22}.

Hedging portfolios that include exotic options is a significant research challenge requiring sequential decision-making to periodically rebalance the portfolio. This has drawn interest in RL-based solutions, which have shown superior performance compared to traditional hedging strategies. For instance, \cite{kolm_ritter} utilized Deep Q-learning and Proximal Policy Optimization (PPO) algorithms to develop a policy for option replication considering market frictions, capable of handling various strike prices. \cite{2021_giurca_delta} employed RL for delta hedging, taking into account transaction costs, option maturity, and hedging frequency, and applied transfer learning to adapt a policy trained on simulated data to real data scenarios. \cite{2q_Cao_2020} and \cite{deeper_hedging_Gao_2023} applied the Deep Deterministic Policy Gradient (DDPG) algorithm to hedge under SABR and Heston volatility models, respectively.

Additionally,~\cite{2022_empirical_deep_hedging} trained an RL algorithm on real intraday options data spanning six years for the $S\&P500$ index. \cite{jpmorgan_market_friction} addressed hedging over-the-counter derivatives using RL under market frictions like trading costs and liquidity constraints. \cite{cva_hedge} explored RL for CVA hedging. The literature extensively covers RL-based sequential decision-making for European~\cite{kolm_ritter,2q_Cao_2020,halperin2019qlbs}, American~\cite{pmlr-v5-li09d_earlyexerciseRL}, and Barrier options~\cite{chen2023hedging_barrier}, demonstrating that RL is a compelling alternative to traditional hedging strategies based on performance and $PnL$ distribution. However, there has been limited focus on hedging high-risk options such as autocallable notes. Recently, \cite{autocall_hedge_mc} introduced a method for pricing Autocallable notes and employing a Delta neutral strategy for hedging, highlighting the complexity due to multiple barriers and the high cost of Delta hedging near these barriers.

Our approach involves training an RL policy to hedge a 4-year maturity autocallable note portfolio with three underlying assets. We evaluate the RL agent's performance under varying transaction costs and with different hedging instruments, demonstrating that the RL agent achieves a better $PnL$ distribution compared to traditional hedging strategies.

\section{Autocallable Note Pricing}
\label{sec:price}

Autocallable notes are structured products that provide the investors with an opportunity to earn extra interest in terms of coupon payments if the underlying asset price closes above a specific threshold on periodic observation dates (barriers). In addition, the note will be autocalled or redeemed on an observation date if asset price return is above or equal to autocall barrier. Otherwise, it may offer contingent downside protection when the notes are held to maturity. 

An autocallable note also provides principal protection at maturity if the Reference Index Return is greater than or equal to -30\% on the final valuation date. Table~\ref{tab:autocall_str} shows the note structure that we used in our experiments. We use our ML pricer to simulate the note price. 
The note greeks ($Delta$ and $Gamma$) are generated using the finite difference method. 
The note greeks change significantly near the observation date, making it difficult to maintain a Delta/Gamma neutral portfolio (as also highlighted in~\cite{autocall_hedge_mc}). 


\begin{table}[t]
\caption{The details of the Autocallable coupon note structure on U.S. Select Regional Banks Index (AR).}
\label{tab:autocall_str}
\centering
 \begin{tabular}{|c | c|}
\hline
 Variable & Value \\ \hline
 Reference Index  & The worst performing of \\&asset A, asset B and asset C \\ \hline
 Initial Price  & A:\$382, B:\$494, C:\$142 \\ \hline
 Term (0\% fee)  & $4$ years \\ \hline
 Coupon Frequency & Quarterly  \\ \hline
 Coupon Rate  & 2.275\%  \\ \hline
 Coupon Barrier  & -25.00\%  \\ \hline
 Autocall Frequency  & Quarterly \\ \hline
 Call Barrier  & 0.00\% \\ \hline
 Contingent Principal & \\ Protection  & -30.00\%  \\ \hline
 \end{tabular}
 \end{table}

Pricing autocallable notes involves intricate considerations of various risk factors, such as underlying assets, interest rates, and volatility. The inclusion of potential coupon and call features complicates the price profile over time (for example, see Figures~\ref{fig:autocall_3_underlying_4yr_monthly_call} for the price grid at different call time). Traditional pricing models like Monte Carlo Simulation are time-consuming, hindering applications like Reinforcement Learning or XVA calculations. Utilizing machine learning (ML) to approximate the original pricing model offers substantial efficiency gains. For instance, in a case of autocallable Notes with three underlying indexes, the ML approximator executes the task 250 times faster than the original pricer. Moreover, the execution time of the ML approximator remains constant regardless of the original pricer's complexity, resulting in significant efficiency enhancements.


In this paper, we apply Chebyshev Tensor~\cite{ML_MC_5_zeron2023frtb,ML_MC_6_maran2021chebyshev,ML_MC_7_ruiz2021machine,ML_MC_8_zeron2021denting} as the model-free machine learning pricing approach. Chebyshev interpolants possess two unique mathematical properties:

\begin{enumerate}
    \item They converge exponentially to analytic functions.
    \item Algorithms exist that ensure a fast and numerically stable evaluation of approximation functions.
\end{enumerate}

Chebyshev points, also known as Chebyshev nodes, are specific points in the interval $[-1,1]$ used in polynomial interpolation. They are particularly important because they minimize the problem of Runge phenomenon, which is the oscillatory behavior seen at the edges of an interval when using high-degree polynomials to approximate functions.

The $n$ Chebyshev points of the first kind are defined as:

\begin{equation}
    x_i = \cos\left(\frac{2i-1}{2n}\pi\right) \quad \textrm{for}\; i=1,2,\dots,n
\end{equation}

These points are the roots of the Chebyshev polynomial of the first kind, $T_n(x) = \cos(n\arccos(x)$

The most efficient way to evaluate Chebyshev interpolants and their derivatives is through the barycentric formula\cite{ML_MC_6_maran2021chebyshev,ML_MC_7_ruiz2021machine}. Let $x_0,\dots,x_n$  be the first $n+1$ Chebyshev points and let $f_0,\dots,f_n$ be the values of $f$ (the original derivative pricing functions) on these points. Then the polynomial interpolant $P_n$ to $f$ on $x_0,\dots,x_n$ is given by

\begin{equation}
\label{eq:cheb_poly_inter}
    P_n(x) = \frac{\sum_{i=0}^{n} \sigma \frac{(-1)^i f_i}{x-x_i}}{\sum_{i=0}^{n} \sigma\frac{(-1)^i}{x-x_i}}
\end{equation}

Where $P_n(x) = f_j$ if $x=x_j$. The symbol $\sigma$ in equation~\ref{eq:cheb_poly_inter} means the summation is multiplied by $0.5$, when $i=0,n$.

For the autocallable note, let the risk factors (underlying asset prices, volatility, and interest rates) be the variables $x$. By taking n Chebyshev points and evaluating the prices $f_0,\dots,f_n$ at these points using the Monte Carlo simulation method, we can train a Chebyshev Tensor via the barycentric method. However, as illustrated in Figure 1, the price of an Autocallable note is not differentiable in the time to maturity dimension due to the autocall feature and potential coupons. Therefore, we cannot use the time to maturity as a variable; instead, we should fix the date and create a Chebyshev Tensor for each day.

\section{Hedging as an MDP}
\label{sec:problem}

In this section, we will describe the problem formulation using Markov Decision Process (MDP) using which we train a reinforcement learning based agent to achieve hedging objective. 

\subsection{Problem Formulation as an MDP}
The RL Pipeline uses a MDP (Markov Decision Process) to frame the problem of our RL agent interacting with the environment to maximize a potential reward. An MDP is defined as a tuple of elements ($S,\mathcal{A},f,R,\gamma$), where $S$ is the state space, $\mathcal{A}$ is the action space, $f(s_t,s_{t+1})$ is the state transition function, $R(s,a)$ is the reward function and $\gamma$ is the discount factor. We formulate a finite horizon discounted sum reward problem where the horizon length is the maturity of the option ($4$ year in our experiments). 

\noindent The individual elements of the MDP are described below: 

\noindent\textbf{State:} 
The state represents the information available to the agent at each time step. In the current hedging framework, the state at time $t$ is given by $s_t = (x_t, \gamma_p, \tau)$, where $x_t$ contains the stock price of all underlying indices/assets, $\gamma_p$ is the portfolio gamma, and $\tau$ indicates the time remaining until the next call date.

\noindent\textbf{Action:} RL hedging agent’s action at any instant is the proportion of maximum hedging that can be done. For example, if the selected action is $0.2$, then the RL agent takes a position equal to the $20\%$ of the maximum hedge allowed. This proportion is then translated into actual number of units in the hedging instrument and those units are used to simulate the portfolio for the next time step. 

\noindent\textbf{Reward:} Reward is defined as the following:

\begin{equation}
\label{eq:reward}
R_i = -\kappa |V_i H_i| + (P_i^- - P_{i-1}^+)
\end{equation}

where $V_i$ is the value of the option (hedging instrument) at time $i\Delta t$, $H_i$ is the position taken in the hedging instrument, $\kappa$ is the transaction cost, $P_i^-$ is portfolio’s market value before time $i\Delta t$, $P_i^+$ is portfolio’s market value after time $i\Delta t$, $-\kappa |V_i H_i|$ is the transaction cost paid at time $i$, and $(P_i^- - P_{i-1}^+)$ is the change in portfolio value from time $(i-1)\Delta t$ to $i\Delta t$.

\noindent\textbf{State transition function:} With $s_t$ as the state at time $t$, the policy selects an action $a_t\in [0,1]$. The next state is updated based on the amount of hedging done at the current instant. The updated gamma of the portfolio and the new underlying price are then used in the state variable at the next instant. 

Using the above MDP, an RL environment was created which gives the next state of the environment based on the actions selected by the RL agent. At any given time, the RL agent interacts with the environment simulator by providing an action and the environment returns the next state and the reward.

\begin{figure}[h]
\centering
\includegraphics[width=8.5cm]{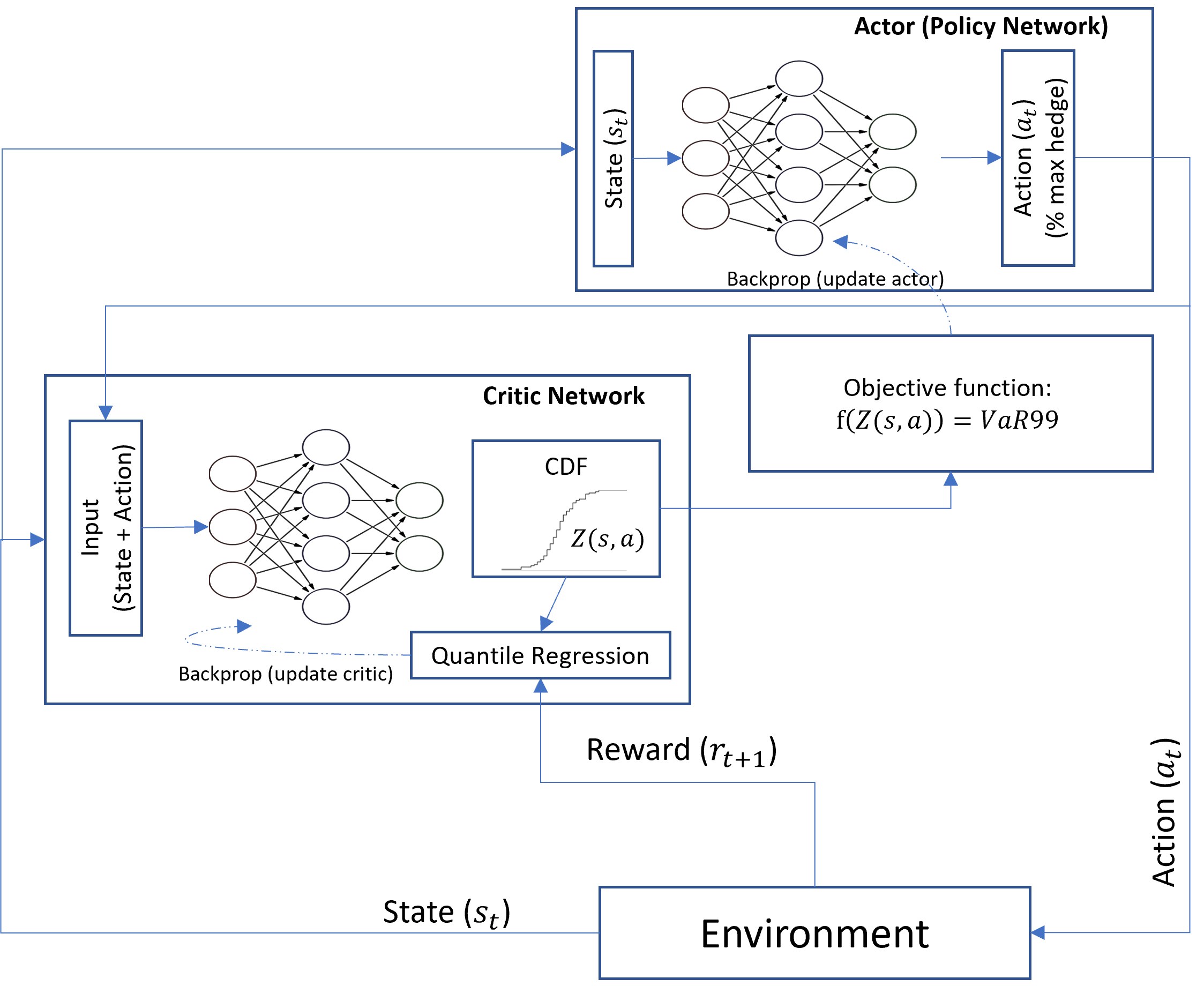}
\caption{Model architecture of Distributed Distributional DDPG (D4PG) with Quantile Regression (QR).}
\label{fig:d4pg}
\end{figure}

\subsection{Distributional Reinforcement Learning}
 We use Distributional Reinforcement Learning (DRL) in our proposal. DRL is an advanced paradigm which focuses on modeling the entire distribution of returns, rather than solely estimating the expected value. In DRL, the return $G_t$ (total reward received by the agent in an episode) is modelled as a distribution $Z_t^\pi$ for a fixed policy $\pi$. The return distribution provides more information and is more robust as compared to only the expectation. Classical RL tries to minimize the error between two expectations, expressed as  
$\E_{s,a,s^{'}} [\{r(s,a) + \gamma max_{a^{'}} Q(s^{'},a^{'}) - Q(s,a)\}^2]$, where $Q(s,a)$ is the output of the policy and $r(s,a) + \gamma max_{a^{'}} Q(s^{'},a^{'})$ is the target function. In contrast, in DRL, the objective is to minimize a distributional error, which is a distance between the two full distributions (the target PMF and the predicted PMF)~\cite{bellemare2017distributional_c51,dabney2017distributional_qrdqn}.

The optimal action is selected from the $Q$ value function:
\begin{equation}
    a^{*} = \argmax_{a^{'}} Q(s^{'},a^{'}) = \argmax_{a^{'}}  \E[Z(s^{'}, a^{'})]  
\end{equation}


In DRL, the distribution of returns is represented as a PMF (Probability Mass Function) and generally, the probabilities are assigned to discrete values that denote the possible outcome of the RL agent. Let’s say we have a neural network that predicts this PMF by taking a state $s$ and returning a distribution $Z(s,a)$ for each action. Categorical distributions are commonly employed to model these distributions in some DRL algorithms like C-51~\cite{bellemare2017distributional_c51} where the action distribution is modelled using a finite number of possible outcomes. In C-51, probabilities are estimated to these fixed locations. We employ Quantile Regression (QR) to learn the distribution of returns and unlike C-51, QR estimates the quantile locations where each quantile corresponds to a fixed uniform probability. That means, QR provides the flexibility to stochastically adjust the quantile locations in place of fixed locations in C-51. QR is a popular approach in DRL which is combined with several distributional RL algorithms such as in QR-DQN~\cite{dabney2017distributional_qrdqn}. QR-DQN uses quantile regression with traditional DQN to learn a distribution of outcomes. 

\begin{figure*}[h!]
    \centering
    \begin{subfigure}[t]{0.3\textwidth}
        \centering
        \includegraphics[width=2.1in]{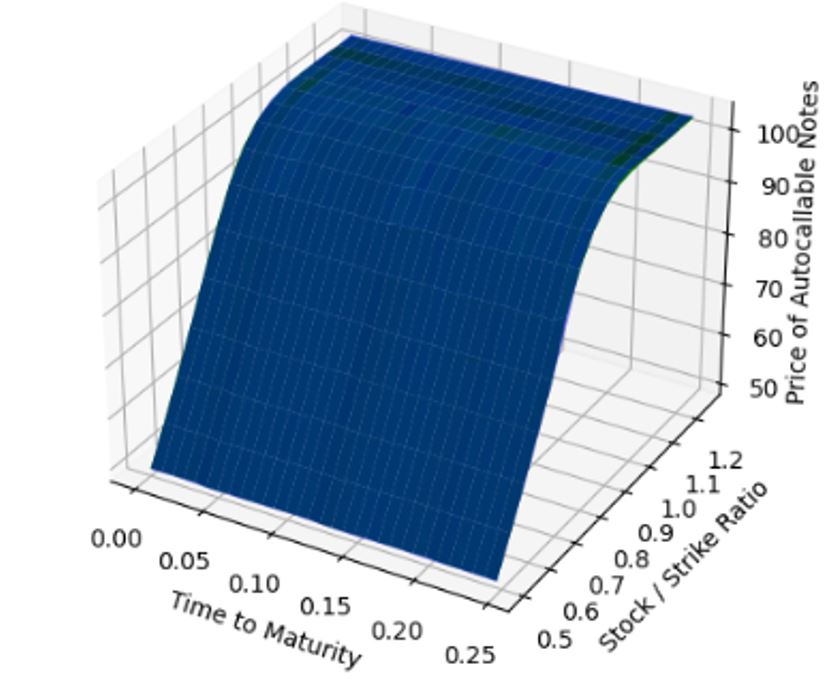}
         \caption{Price grid of ML Approximator (three underlying)} \label{fig:price_grid_3_underlying_ML}
    \end{subfigure}%
    ~ 
    \begin{subfigure}[t]{0.3\textwidth}
        \centering
     \includegraphics[width=2.1in]{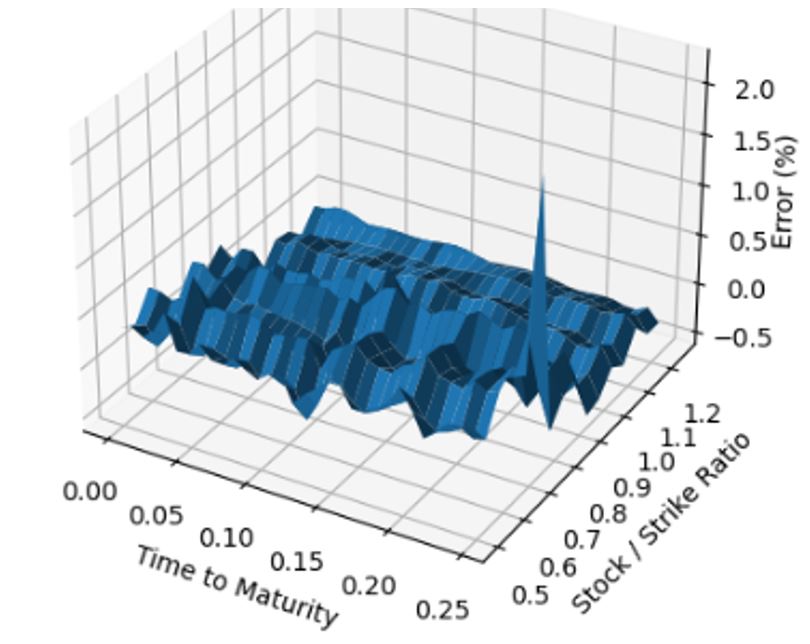}
         \caption{Error(\%) of ML Approximator vs. Monte Carlo pricer}
         \label{fig:error_percentage_ML_MC}
    \end{subfigure}
    ~
    \begin{subfigure}[t]{0.3\textwidth}
        \centering
     \includegraphics[height=1.45in]{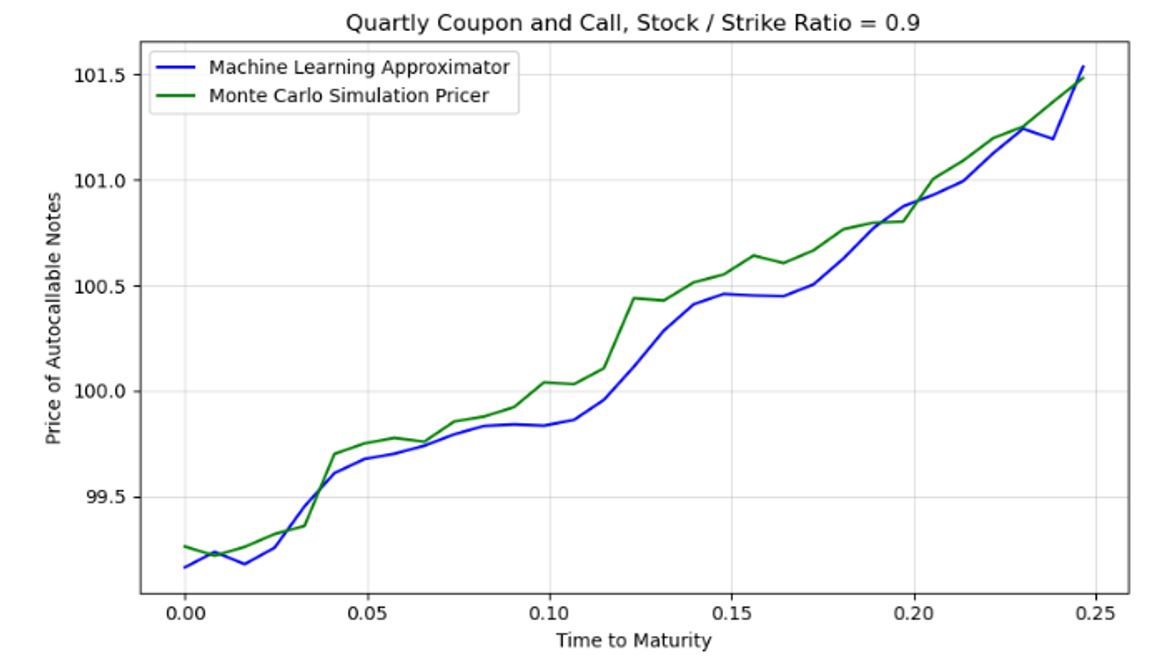}
         \caption{Price comparison of ML approximator and Monte Carlo Simulation}
         \label{fig:price_comparison_ML_MC}
    \end{subfigure}
    \caption{Figure shows the price grid of the autocallable note with three underlying assets and the price comparison with Monte Carlo method. }
\end{figure*}

\subsubsection{Distributed Distributional DDPG (D4PG)}
We use the Deep Deterministic Policy Gradients (DDPG) algorithm~\cite{d4pg_orig} to learn the underlying distribution of returns and the optimal policy for hedging action selection. DDPG is an Actor-Critic based method which helps to learn a policy in continuous action space. 
D4PG is a distributional RL algorithm which estimates the distribution of the return (unlike the mean in classical RL). Figure~\ref{fig:d4pg} shows the model architecture that we utilized in our experiments to train the D4PG algorithm. 
There are three components:
\begin{enumerate}
    \item \textbf{The trading environment}: One essential component of the architecture is the trading environment, which simulates the stock and option prices. It tracks how the portfolio evolves over time based on the agent's hedging position, market dynamics (e.g., SABR volatility), and the arrival of the client options. At any particular time, the environment receives the hedging action and returns the next state and the reward.
    \item \textbf{The actor neural network} (also known as policy network) implements the hedging strategy. It is a neural network of size $(256, 256, 256, 1)$. 
    At any instant $t$, it takes as input a state $s_t$ and outputs the amount of hedging ($a_t$) that the agent should perform. The objective function for training the agent's neural network is $99\% VaR$. 
    \item \textbf{The critic neural network} takes as inputs a state, $s_t$, and the action from the actor’s output, $a_t$. Its role is to (a) estimate the distribution of the trading loss at the end of the hedging period, $Z(s_t, a_t)$, when taking action $a_t$ in state $s_t$, and (b) compute gradients that minimize the objective function $f(Z(s_t, a_t))$. We use a neural network of size $(512, 512, 256, 1)$ as the critic network. We use the reward from the environment to train the agent’s critic network. 

\end{enumerate}

We utilize quantile regression (QR)~\cite{dabney2017distributional_qrdqn} in combination with D4PG to approximate the distribution $Z(s, a)$ with the help of quantiles at the output of critic neural network. We use $100$ quantiles in our experiments for D4PG policy learning. Each quantile has a fixed probability but the location is stochastically adjusted during training. 
The policy output and the target distribution are used to compute the error in the policy output. Wasserstein distance~\cite{dabney2017distributional_qrdqn} is used as the loss function to estimate the quantiles which is later used to backpropagate the gradients using \textit{Adam} optimizer. 

\section{Experiments and Results}
\label{sec:results}

In this section, we will describe the different empirical analysis that we performed to get the pricing faster, hedging on autocallable notes with three underlying assets, along with the experimental setup and the performance metrics. 

\begin{figure}[h]
\includegraphics[width=9.5cm]{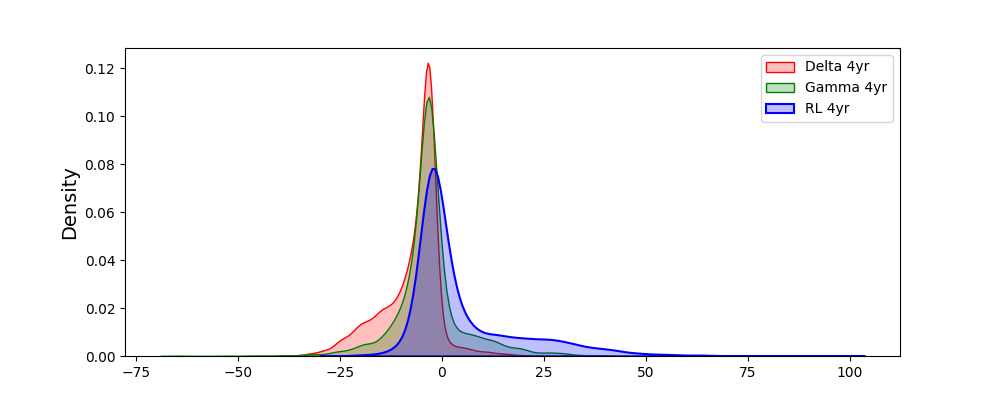}
\caption{The $PnL$ distribution of the traditional Delta neutral and Delta-Gamma neutral hedging strategies and the $PnL$ of RL agent based hedging strategy on a portfolio containing autocallable note with $4$ year maturity with three underlying assets.}
\label{fig:perf_comparison_american}
\end{figure}

\begin{table*}[h]
\caption{The table compares the performance of the Delta neutral, Delta-Gamma neutral, and RL agent on a portfolio of autocallable note (having three underlying assets) with American options (Put and Call) as hedging instruments with $2\%$ transaction cost. \\Note: $VaR$ numbers are actual $PnL$ and not the opposite side.}
\label{tab:perf_comp}
\centering
 \begin{tabular}{|c c c c c c c c c|}
\hline
 Strategy & 99\%VaR & 99\%CVaR & 95\%VaR & 95\%CVaR & 5\%$VaR$ & 5\%$CVaR$ & 1\%VaR & 1\%CVaR \\ \hline
 Delta Neutral  & -28.1 & -30.82 & -21.62 & -25.55 & -0.04 & -7.87 & 10.82 & -7.38  \\
 Delta-Gamma Neutral & -26.05 & -33.59 & -15.38 & -22.58 & 13.05 & -4.22 & 26.78 & -3.32  \\
 RL [Am. Put \& Call] & -11.65 & -15.46 & -6.35 & -9.67 & 33.95 & 2.76 & 50.36 & 4.27  \\


 
 \hline
     
 \end{tabular}
\end{table*}

\subsection{Experimental Setup}
We are hedging a trader's portfolio which contains risky options - the trader hedges the portfolio by adding other instruments to a hedging portfolio at every hedging instant. In other words, the trader rebalances the portfolio at every $\Delta t$ time interval for hedging. 
The hedging action involves selecting the percentage of $Gamma$ to hedge at any rebalancing instant.
For example, in the experiments for autocallable note hedging, the trader's portfolio contains one short Autocallable note with $4$ years of maturity and featuring three underlying indices and $2\%$ transaction cost. The trader also keeps a hedging portfolio and 
at every hedging instant ($\Delta t = 1$ month), the trader adds an at-the-money American call and put option to the hedging portfolio for hedging. A hedging strategy decides how much hedging needs to be performed based on the different portfolio Greeks such as $Delta$, $Gamma$, etc. $Delta$ tells how much the option's price will change for a one-point change in the underlying asset's price, while $Gamma$ tells how much the $Delta$ will change for a one-point change in the underlying asset's price. Trader's use these Greeks to manage risk and make informed decisions. For example, a Delta neutral strategies hedges the entire $Delta$ of the portfolio by creating a position with a $Delta$ value of zero, or very close to zero. 
We propose a RL based hedging strategy which learns to decide the percentage of the $Gamma$ that needs to be hedged. We compare the hedging performance of our RL agent with traditional hedging strategies, Delta neutral and Delta-Gamma neutral. 

Our RL agent uses the D4PG (Distributed Distributional DDPG) algorithm with quantile regression to learn an optimal policy. The RL agent selects an optimal action using the learned policy from interval $[0,1]$. The RL agent is trained for $40,000$ episodes (where one episode is one stock path till note maturity or auto-call). To compare the performance for all methods, we generate an additional $5000$ episodes on which KPI metrics are reported. The prices for the underlying asset are generated using Geometric Brownian motion (GBM) model. We evaluate the hedging performance using $95\%$ percentile of the value-at-risk ($VaR$) and conditional value-at-risk ($CVaR$) distribution. 
 
\noindent\textbf{Implementation:} We implemented the D4PG algorithm using the ACME library from Deepmind~\cite{hoffman2020acme} with Tensorflow backend. We utilized a server with $64$-GBs of RAM with $16$-CPU cores for training. To train the model, we employ a deep neural network with three hidden layers with \emph{Adam} optimizer for both actor and critic networks. The subsequent sections present the performance on the optimal parameters.

 
     

\begin{table}[t]
\caption{Runtime comparison: Machine Learning Approximator VS. Monte Carlo Simulation based Pricer (over 100,000 paths).}
\label{tab:autocall_price_comp}
\centering
 \begin{tabular}{|c | c | c | c|}
\hline
 Instrument & MC Pricer& ML  &  Efficiency\\
 & & &Gain\\ \hline
 Autocall Note &&& \\(7 years, &&& \\$1$ Underlying Asset)  & 1.87 sec & 1e-5 sec & 190,000x \\ \hline
 
 Autocall Note &&&\\(4 years, &&&\\$3$ Underlying Asset)  & 1.25 sec & 5e-3 sec & 250x \\ \hline
 
 \end{tabular}
 \end{table}
 
\subsection{Note Pricing}

Figure~\ref{fig:price_grid_3_underlying_ML} illustrates the price grid for the first three months of an autocallable note with a four-year maturity, using multiple underlying assets. This figure highlights how the prices evolve over time for the given period. In figure~\ref{fig:price_comparison_ML_MC}, we present a comparison of the price estimations obtained using machine learning (ML) approximation (Chebyshev Tensor) and the traditional Monte Carlo (MC) simulation method. The maximum absolute error that we encountered in our experiments is $0.43\%$ between both the pricers. This comparison shows that the ML method can approximate the original pricing function (MC) very well. There is no significant bias in their errors. Figure~\ref{fig:error_percentage_ML_MC} depicts the error percentage in pricing between the ML approximation and the Monte Carlo method, specifically for an autocallable note with three underlying assets. This figure quantifies the pricing discrepancies and provides insights into the performance of the ML approximation relative to the Monte Carlo simulation.


 Table~\ref{tab:autocall_price_comp} provides  a detailed comparison of computation times for pricing a four-year autocallable note with three underlying assets. This comparison is essential for understanding how traditional methods like Monte Carlo simulations, which can be computationally intensive, stack up against more modern approaches such as machine learning-based models that promise faster computation. By evaluating these trade-offs on accuracy and computational efficiency, practitioners can make informed decisions on selecting the appropriate pricing method based on their needs for accuracy, computational resources, and operational efficiency.

\subsection{RL Agent for Hedging }
In this experiment, we hedge a trader's portfolio containing one short autocallable note with a maturity of 4 years and multiple underlying assets. We use American call and put options as hedging instrument which are added at every hedging instant based on the action selected by the RL agent. 
For this experiment, the state at any time instant $t$ contains the stock price ($x_t$, portfolio gamma, and the days to the next call date). The trader rebalances the portfolio every month by adding one American option to the hedging portfolio, both call and put option are added to eliminate the need to rebalance the $Delta$ exposure in the portfolio. We choose $\Delta t=1$ month as the rebalancing interval because of the longer maturity time of 4 years for the autocallable note. The hedging agent strategy decides how much hedging needs to be performed in terms of the Gamma value of the portfolio. 
We use $VaR 99\%$ as the objective function as the reward to the RL agent which then selects an action for the amount of hedging that needs to be performed. The RL agent generated the best payoff as compared to the traditional hedging strategies (Delta-neutral and Delta-Gamma neutral). The figure~\ref{fig:perf_comparison_american} shows the $PnL$ distribution of the three hedging strategies and RL agent is found to be performing best than the other strategies. Table~\ref{tab:perf_comp} shows the $VaR-5\%$, $CVaR-5\%$, $VaR-95\%$, $CVaR-95\%$ of the $PnL$ distribution to compare the values at different percentiles of the $PnL$ distribution. Please note that the values in the table are for the actual $PnL$ distribution. The $PnL$ for $VaR-5\%$ of the RL agent is $33.95$, significantly outperforming both the Delta-neutral strategy, which has a $VaR$ of $-0.04$, and the Delta-Gamma-neutral strategy, which has a $VaR$ of $13.05$. Additionally, the $VaR-95\%$ of the RL agent is $-6.35$, which is also better compared to the Delta-neutral and Delta-Gamma-neutral strategies. Overall, the RL agent shifts the $PnL$ distribution positively. From the table and the figure, we can claim that the RL agent provides substantial improvement in the $PnL$ by hedging of the portfolio. 


\section{Conclusion}
\label{sec:conl}

We demonstrated that pricing autocallable structured notes using traditional Monte Carlo simulations is computationally expensive. To address this, we proposed a machine learning option pricing method with Chebyshev tensors that achieves pricing $250$ times faster. Additionally, we introduced a distributional reinforcement learning (RL) method for hedging portfolios of structured products, which significantly improves PnL compared to traditional Delta-neutral and Delta-Gamma strategies. Our RL agent achieves a $VaR$ of $33.95$ at $5\%$, while Delta-neutral and Delta-Gamma strategies yield $-0.04$ and $13.05$, respectively, highlighting the RL agent's superiority.





\bibliographystyle{ACM-Reference-Format}
\bibliography{paper}

\end{document}